\documentclass[amsmath,showkeys]{revtex4}
\usepackage{graphics,color}

\begin{document}

\hfill Published in Chaos \textbf{14}, 72 (2004)

\title{Universality in active chaos}

\author{Tam\'as T\'el}
\affiliation{Institute for Theoretical Physics, E\"otv\"os University, 
P.O. Box 32, H-1518, Budapest, Hungary}

\author{Takashi Nishikawa}
\email{tnishi@smu.edu}
\affiliation{Department of Mathematics, Southern Methodist University, Dallas, TX 75275, USA}

\author{Adilson E. Motter}
\affiliation{Max Planck Institute for the Physics of Complex Systems,
N\"othnitzer Strasse 38, 01187 Dresden, Germany
}

\author{Celso Grebogi}
\affiliation{Instituto de F\'{\i}sica, Universidade de S\~ao Paulo, 
Caixa Postal 66318, 05315-970, S\~ao Paulo, SP, Brazil}

\author{Zolt\'an Toroczkai}
\affiliation{Complex Systems Group, Theoretical Division,
Los Alamos National Laboratory, MS B213, Los Alamos, NM, 87545, USA}

\keywords{reaction, Lagrangian chaos, filamental fractal, inertial effect}

\begin{abstract}
Many examples of chemical and biological processes take place in
large-scale environmental flows.  Such flows generate filamental
patterns which are often fractal due to the presence of chaos in the
underlying advection dynamics.  In such processes, hydrodynamical
stirring strongly couples into the reactivity of the advected species
and might thus make the traditional treatment of the problem through
partial differential equations difficult.  Here we present a simple approach for the activity in in-homogeneously stirred flows.  We show that the fractal patterns serving as skeletons and catalysts lead to a rate equation with a universal form that is independent of the flow, of the particle properties, and of the details of the active process.  
One aspect of the universality of our appraoch is that it also
applies to reactions among particles of finite size (so-called
inertial particles).
%Our technique also applies when the active particles are of
%finite size (so-called inertial particles), and when the filamental
%patterns do not form a fractal in the traditional sense of fractality.
\end{abstract}

\maketitle

{\bf
Environmental processes of biological and chemical nature, like the
plankton blooming in the oceans\cite{Abrah1,Abrah2} and the ozone
depletion in the stratosphere\cite{Edouard,Wonhas}, occur within fluid
flows.  The study of such processes is of importance in a broad range
of fields including chemistry\cite{metcalfe,Epstein,Pai,menzinger,Neu6}, population
dynamics\cite{Martin}, geophysics, atmospheric
sciences\cite{Haynes,Tan,abraham} and combustion\cite{kiss}.  Many chemical and biological species are
immersed in a dynamic environment typically characterized by a
time-dependent flow which advects and stirs the species.  Because the
advection dynamics is often (Lagrangian) chaotic, the
application of the theory of dynamical systems to hydrodynamical
advection problems \cite{AREF,OTTK,Wigg} sheds new light on the
reactive dynamics and on the production efficiency in such flows.  As
a result of the chaotic dynamics, fractal patterns are present, and the product distribution of the reactive process becomes concentrated along these patterns.  There is evidence that such a filamental structure is indeed present in the distribution of active
species in oceanic and atmospheric flows, such as the one shown in
Fig.~\ref{fig:real}.  In this article, we develop a description for active processes in flows with such structure.
\begin{figure}
\begin{center}
  \framebox[3in]{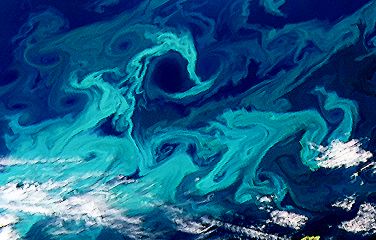 \rule[-.25in]{0in}{.5in}}
\end{center}
\caption{Filamentation in a phytoplankton bloom in Norwegian Sea. (Provided by the \mbox{SeaWiFS} Project, NASA/Goddard Space Flight Center, and ORBIMAGE, URL: {\tt http://visibleearth.nasa.gov/cgi-bin/viewrecord?5278}) \label{fig:real}}
\end{figure}  
%% We argue that this approach enjoys much broader applicability than was considered before.  
Our results imply that the character of a reaction can drastically change if it takes place in a time-dependent flow. A reaction which spreads over the whole space in a well mixed container can lead to a pattern formation of a new type: the product is asymptotically distributed around a filamental fractal which moves in a rhythm corresponding to the time dependence of the flow. In a periodic case, the total amount of product is thus oscillating around a mean: a kind of limit-cycle behavior sets in. This pattern formation is due to the interplay of the chaotic particle motion produced by hydrodynamics and the production of the new particles by the reaction.
In particular, we show that the theory is also valid if the reaction takes place with inertial particles, i.e., with particles of small but finite size whose density differs from that of the surrounding fluid.
}

A standard approach to describe active processes in flows is
based on the use of advection-reaction-diffusion equations. The basic required
assumption, from the Eulerian hydrodynamical point of view,
is that {\em the active species behave as fluid particles}, and as
such, their distribution can faithfully be described by smooth
concentration fields.  As a simple example, consider an auto-catalytic
reaction $\mbox{A} + \mbox{B} \rightarrow 2 \mbox{B}$ in which an
unlimited amount of component A is present.  In a homogeneously
mixed environment, the reaction equation is $d{b}/d t= k a b$,
where $a$ and $b$ stand for the respective A and B concentrations and
$k$ is the reaction rate.  If the concentrations of A and B are not
homogeneous in space, diffusion plays a role, and the reaction
outcome is described by a partial differential equation:
\begin{equation}
\frac{\partial b}{\partial t} +
({\bf u} \cdot \nabla) b = k a b+ \kappa \Delta b,
\label{eqn:ard}
\end{equation}
where $\kappa$ is the diffusion coefficient and
$\bf u$ is the velocity of the fluid (and $a$ is the fixed
concentration of the component A).  This is the
advection-reaction-diffusion equation, in which the presence of the
advection term $({\bf u} \cdot \nabla) b$ indicates that, in this
approach, the advection of particles must take part with the fluid
velocity $\bf u$.  Although this approach is capable of describing a
filamentation process\cite{Neu1, Neu2, Neu4, Neu5}, it might break
down if strong clusterization takes place \cite{Sta1,Sta2} as in the
case where inertial effects of the advected reacting
particles are included.

Inertial effects are due to the fact that the active particles are
{\em of finite size}, and they can be {\em heavier or lighter} than
the fluid.  This is the case with aerosol particles or cloud rain
droplets\cite{Falkovich} in the air (heavier) and with gaseous
bubbles in fluids or some species of phytoplankton in oceans (lighter).  Such
particles, due to viscosity (Stokes drag), try to follow the
surrounding fluid, but typically diverge from the fluid
trajectory\cite{Max1, Max2, Yu, Cris,
Tanga, Elperin, Bracco, Babiano, Shinbro, Cart, Benc}.  The inertial effect alone is a source of chaotic
behavior\cite{Yu, Nishi1, Nishi2, Liu}.  In general, because of
viscosity, the dynamics of inertial particles is dissipative and it is
characterized by the presence of attractors.  An attractor embodies
the general tendency towards accumulation or clusterization of
inertial particles (see Appendix \ref{appx:A}).  As a result, the advection-reaction-diffusion
equation~(\ref{eqn:ard}) {\em is no longer valid} for the inertial
active particles since the particle
motion differs from that of the surrounding fluid.  A partial
description in terms of concentration is available, but only for the
continuity equation and the diffusion equation\cite{Balk, Lopez}, or in the limit of small inertia\cite{Elperin,Reigada},
which can only be a starting point for more general problems.  At
present, it is unknown whether an equation, analogous to
Eq.~(\ref{eqn:ard}), exists at all for active inertial particles.
This fact alone calls for an alternative description of the global
kinetics of the active species.

Herewith we present an approach that can successfully be applied to
describe the kinetics of active particles in a flow.  The idea is to
determine the total number of
particles of a given constituent in a macroscopic range of the fluid
after some time, and show that this quantity fulfills a kind of
simple rate equation.

The essential ingredient of our theory is the existence of a fractal
dimension $D$ in the physical space of the fluid for the reaction-free
advection problem.  This fractality means that, after some transient
time (reacting or non-reacting) particles accumulate along zero-width filaments
of a $D$-dimensional fractal set in the fluid.  There are some pieces of evidence that suggests the presence of such filamental fractal sets in a variety of processes taking place in flows\cite{Abrah1,Abrah2,Edouard,Wonhas,metcalfe,Epstein,Pai,menzinger,Neu6,Martin,Haynes,Tan,abraham,kiss}. There are two distinct dynamical origins for such fractal sets.  The better known of them is the
presence of a {\em chaotic attractor}\cite{Yu, Babiano, Shinbro,
Cart, Benc, Nishi1, Nishi2, Liu} in the inertial problems.  The other
possibility is for the advection dynamics to exhibit {\em transient
chaos}\cite{Motter1, Motter2, Toro, Karo}, which can be associated
with chaotic saddles in the advection dynamics of either the inertial or noninertial problems.  The fractal filaments align along the unstable
direction, where stretching takes place, and across these filaments,
contraction occurs. In typical chaotic systems, this contraction is
exponential in time with some contraction rate $\lambda>0$. In the
language of dynamical system theory, $-\lambda$ is the largest among the
contracting Lyapunov exponents of the advection dynamics which, of
course, depends on the flow's hydrodynamical characteristics and the
particle's inertial properties.  This means that a typical distance of $\delta$ across the filaments shrinks with a velocity $-\lambda \delta$.

In what follows we restrict our attention to autocatalytic reactions
which occur often in nature\cite{Epstein}.  For such processes,
the reaction typically propagates in the form of fronts
(the stable B phase propagates into the unstable A phase) with
relatively sharp boundaries, since the effect of diffusion is rather weak on the
length scales of interest\cite{Neu3}.  In the simplest approximation,
both ozone depletion and plankton blooming can be described by front
propagation of this type.  Further examples are the
Belousov-Zhabotinskii reaction\cite{zaikin} and the propagation of flames\cite{flame}.

Many real examples of environmental flows are essentially
two dimensional, and thus we restrict ourselves here to the case where
$D$ is between 1 and 2.  The flow has a clean fractal structure when
$D$ is well defined and it is different from both 1 and 2.  We note, however, that a similar treatment is applicable even to a three-dimensional flow with fractal invariant manifolds, since the arguments below are quite general.

\begin{figure}
\begin{center}
  \includegraphics{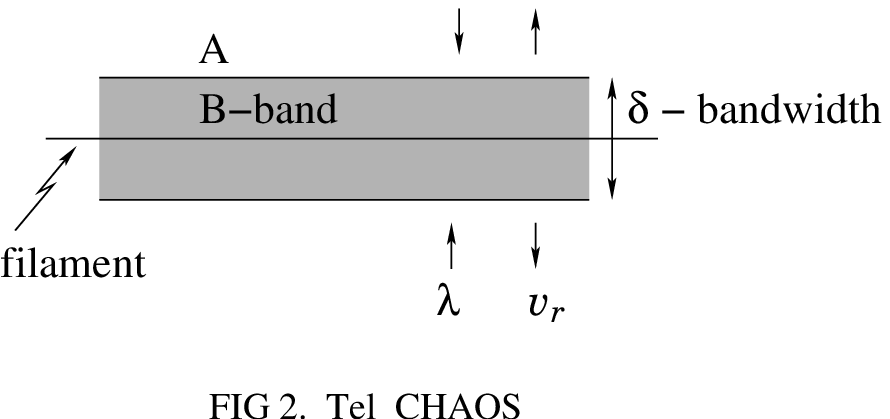}
\end{center}
\caption{Local reaction-advection dynamics on a filamental segment. \label{fig:react}}
\end{figure}

Consider a blob of B-particles in a region of interest.  After some time, material B will be distributed along filaments in
bands of average width $\delta$.  Consider now a {\em single}
filamental segment on the underlying fractal, covered by a band of
B-particles with the average width of $\delta$, as illustrated in
Fig.~\ref{fig:react}.  
The width of the band decreases at the rate $-\lambda \delta$ but
there is, due to the reaction, also an increase of it.  Approximating the
local reaction front velocity by its average value $v_r$, the rate of
increase of the width is $2 v_r$.  Thus, the time derivative
$\dot{\delta} \equiv d\delta/dt$ of the width is given as
\begin{equation}
\dot{\delta}= -\lambda \delta + 2 v_r.
\label{de}
\end{equation}
This simple differential equation expresses the competition of two
effects: the exponential contraction towards the underlying fractal, and the
linear expansion due the autocatalytic process, which could be regarded as a kind of ``infection''.  After some transient time, this competition leads to a
steady state ($\dot{\delta} = 0$) characterized by the fixed point value
$\delta^*=2 v_r/\lambda$.

\begin{figure}
\begin{center}
  %\resizebox{\textwidth}{!}{\includegraphics{fig3.eps}}
  \framebox[3in]{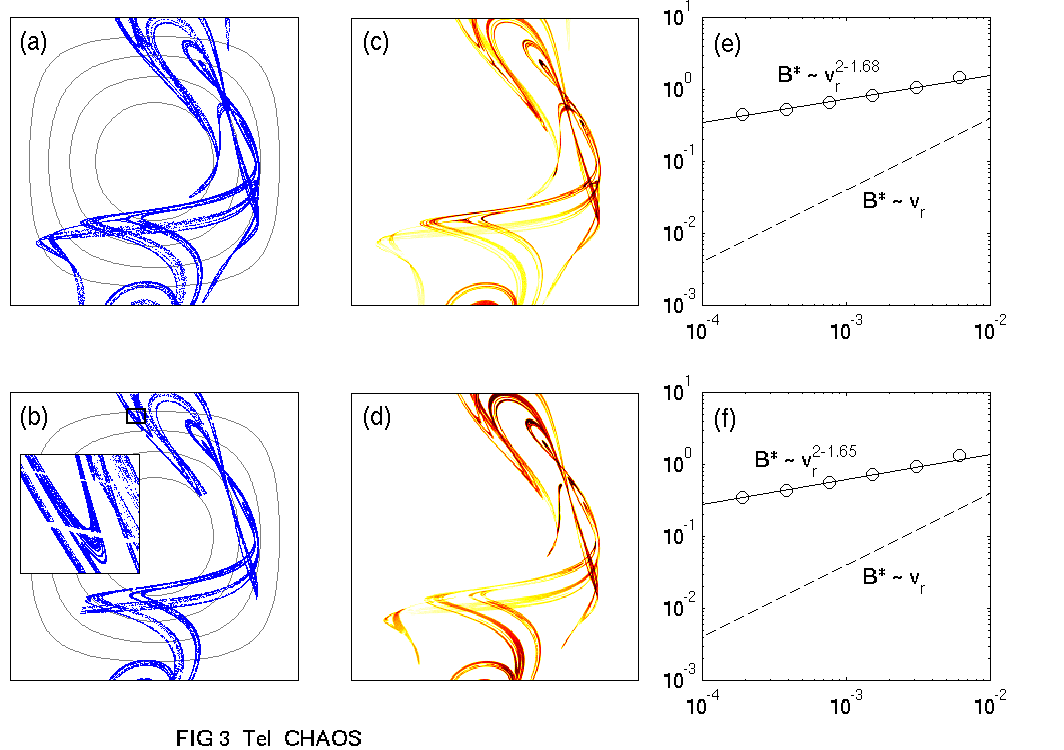 \rule[-.25in]{0in}{.5in}}
\end{center}
\caption{Autocatalytic inertial particles advected by a
two-dimensional flow field of counter-rotating array of vortices with
time-periodic intensity\cite{Yu}, given by the stream function \eqref{stream} with $\omega = \pi$.  The flow
can be regarded as a model of the arrangement of eddies in a
vertical plane of the lower atmosphere or the upper ocean.  The
parameter $k$ measures the amplitude of oscillation of the strength of
the vortices.  (a), (b) Chaotic sets in the advection dynamics of the
nonreacting inertial B-particles: (a) chaotic attractor ($k = 0.53$)
and (b) chaotic saddle responsible for transient chaos ($k = 0.524$).  
The small rectangle at the top of the panel (b) is magnified in the inset to show the small scale (Cantor-like) structure of the saddle.  The other parameters are $St = 1$, $\alpha = 1.7$, and $w = -3.934$
(see Appendix \ref{appx:A}).  (c), (d) Distribution of the product particles after
a sufficiently long time ($t = 100$ periods of the flow field).  Fractal filamentation is caused by the chaotic attractor in (c) and by the chaotic saddle in (d).
Only a single vortex cell of  $[0,\pi]\times[0,\pi]$ is shown and the gravity points downward in (a)-(d).  The color coding
represents the density of B-particles, in which darker colors
correspond to higher density.  The density of B-particle is bounded in this simulation\cite{Nishi1}.  Note that in (d) the product is
distributed not along the chaotic saddle but along its unstable
manifold [the difference between the saddle in (b) and the
unstable manifold seen in (d) is apparently small, partly because they are
projected onto the two-dimensional configuration space].  The initial
condition is a small blob of B-particles.  (e), (f) The total number
$B^*$ of the B-particles in the steady state vs the reaction velocity
$v_r$ in the case of the chaotic attractor in (e) and the
chaotic saddle in (f).  The full line corresponds
to the fit $B^* \sim v_r^{2-D}$ predicted by the theory [see Eq.~\eqref{eqn:bs}].  For
reference, the $B^* \sim v_r$ line is shown as a dashed line, which corresponds to
a nonchaotic case ($D = 1$). \label{fig:sim}}
\end{figure}

The basic observation is that on any filamental segment accumulates
infinitely many other segments, since they form a fractal set.
However, after the spreading of the material along any one of these
fractal segments, there is a fattening up of the segments and we can
assume that these bands are similar to the typical one just been
treated above and have bandwidth $\delta$.  Then, the number of B-particle bands {\em
observed} to cover the segments with an average instantaneous width
$\delta$ is finite.  The union of all bands of B-particles, covering
the fractal filaments, appears to be a fractal on length scales above
$\delta$, but it is a two-dimensional object below
this crossover scale.  Let us consider a fixed region that contains
the filamental fractal bands.  According to
the fractal geometry\cite{Mand,Falc}, the minimal number $N$ of boxes
needed to cover a fractal set of dimension $D$ with boxes of linear
size $\varepsilon$ is proportional to $\varepsilon^{-D}$.  By using the
actual width $\delta$ of the B-coverage as the box size ($\varepsilon =
\delta$), the number of boxes needed to cover the fractal filaments in
the region of observation is $ N{(\delta)} \sim \delta^{-D}$.  The
symbol $\sim$ indicates the presence of a proportionality geometric factor not
written out explicitly.  The total area covered by the B-particles is
therefore $N(\delta)$ times the area $\delta^2$ of a single box, or
$\delta^2 N(\delta)$, which is proportional to $\delta^{2-D}$.  Although the number of B-particles in each box may vary along the
filaments due to the stretching and folding action of the advection
dynamics, the average number of particles in each box will saturate, since the number of A-particles available in each box is limited.  It is then
natural to assume that the area $\delta^{2-D}$ is proportional to the
{\em number} $B$ of the B-particles in a given region of observation,
i.e., $B \sim \delta^{2-D}$.  The time derivative of the total number of
B-particles is $\dot{B} \sim (2-D) \delta^{1-D} \dot{\delta}$, where
$\dot{\delta}$ can be obtained simply by substituting it from
(\ref{de}).  Thus, $\dot{B}$ can be written as the sum of a (negative)
loss term $-L$ and a (positive) production term $P$:
\begin{equation}
\dot{B} = P(B) - L(B),
\label{eqn:db}
\end{equation}
where
\begin{equation}
L(B)= \lambda (2-D) B, \;\;\;\;
P(B)= c v_r (2-D) B^{-\beta}
\label{eqn:cp}
\end{equation}
with $c$ as a $B$-independent geometric factor (which might depend on the
location and the size of the region of observation), and
\begin{equation}
\beta \equiv \frac{D-1}{2-D}.
\label{beta}
\end{equation}
The exponent of the production term is always negative since $1 < D <
2$ ($\beta$ is positive).  Thus, the overall structure of the rate equation is
$$ \dot{B} = -c_1 B + c_2 v_r B^{-\beta}, $$
where $c_1$ and $c_2$ are positive coefficients independent of the reactive process.  The novel feature of this equation is
the singularity of the term $P(B)$.  It states that the smaller is
the number of B-particles, the higher is the production.  This peculiar
scaling property of the production term has been verified in numerical
simulations for autocatalytic noninertial particles\cite{Toro,Karo,Neu3}.  Here we emphasize that these results are {\em
valid for inertial particles} as well.  Figure~\ref{fig:sim} shows the
results of the numerical simulations for finite-size active B
particles in a simple two-dimensional cellular flow field, given by the stream
function \begin{equation}
  \psi(x,y)=[1+k \sin{(\omega t)}] 
  \sin x \sin y, \label{stream}
\end{equation}
where $k$ and $\omega$ are the amplitude and angular frequency of the temporal
oscillation of the flow field, respectively\cite{Nishi1,Nishi2,Liu}.

\begin{figure}
\begin{center}
  \includegraphics{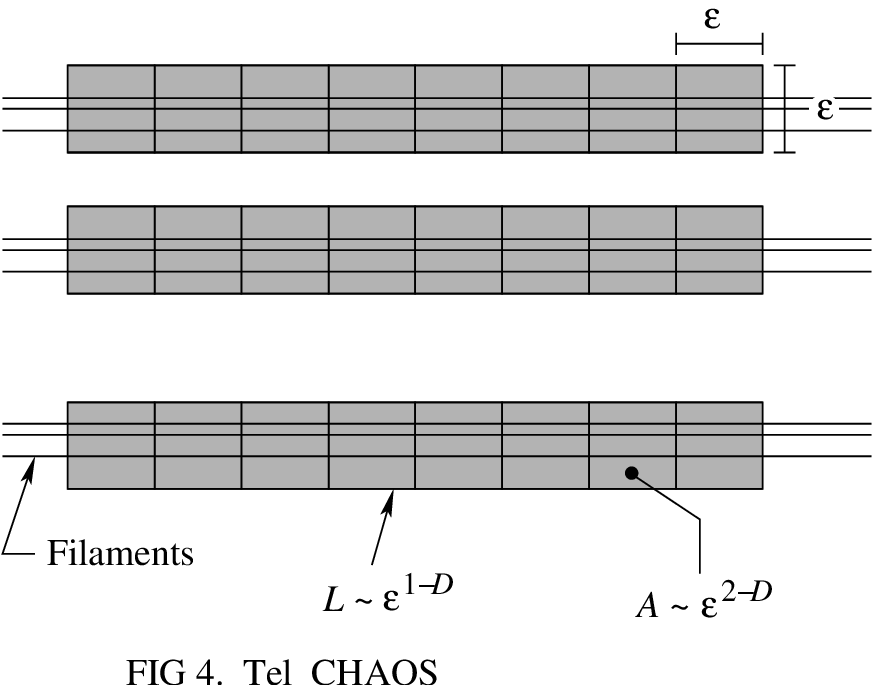}
\end{center}
\caption{Schematic diagram illustrating the
$\varepsilon$-dependence of the perimeter length ${\cal L}$ and the area
${\cal A}$ of a filamental fractal, when observed with resolution
$\varepsilon$. \label{fig:cover}}
\end{figure}

The universality of our description is grounded on the generic
property of a filamental fractal that the perimeter length of its
finite-width coverage {\em increases} as the area of coverage
decreases.  This relationship leads to singularly enhanced reactivity.
In order to see this, let us derive the relationship between the
observed perimeter length ${\cal L}$ and the area ${\cal A}$ of
filamental fractals.  Since the covering of such a fractal set with
small squares of linear size $\varepsilon$ requires $N(\varepsilon)
\sim \varepsilon^{-D}$ such squares, and since two of the four edges
of each box typically belong to the perimeter of the coverage, the
perimeter length is proportional to $\varepsilon^{1-D}$ and {\em
increases} with refining resolution ($D > 1$) (see
Fig.~\ref{fig:cover}).  
On the other hand, the area is proportional to $\varepsilon^{2-D}$ and
{\em decreases} with refining resolution ($D < 2$).  By eliminating
$\varepsilon$ from the relations ${\cal L} \sim \varepsilon^{1-D}$ and
${\cal A} \sim \varepsilon^{2-D}$, we find that
\begin{equation}
{\cal L} \sim {\cal A}^{-\beta}
\label{eqn:la}
\end{equation}
with $\beta$ as given by (\ref{beta}).  Thus, the perimeter length is,
at any small resolution, a {\em negative} power $(- \beta)$ of the
area
\cite{footnote}.  (Note that for
classical nonfractal objects, e.g., a sphere or a cube, one has
${\cal L} \sim {\cal A}^{1/2}$, which is non-singular since the
exponent is positive.)  In view of this, the production term $P$
in~(\ref{eqn:db}) can be interpreted as the expression for the fact
that the reaction takes place along the perimeter of the fattened-up
filamental fractal seen at resolution $\delta$.  Since the peculiar
relationship~(\ref{eqn:la}) is purely geometrical, it does
not depend on the precise nature of the activity.  Similar singular terms appear in the equation for other types of
activity as well\cite{Toro,Karo}.

In summary, the fractal filaments of the advection problem act as
dynamical {\em catalysts} for the reactions.  The rate equation
(\ref{eqn:db}) has a {\em universal} character, as its form does not
depend either on the particle, flow or reactivity properties.
Fundamentally, the exponent $\beta$ characterizes the geometry of the
reaction-free chaotic advection.  The singular productivity disappears for
$D=1$, representing a flow in which the filaments do not form a
fractal, and the advection is consequently nonchaotic.

Equation (\ref{eqn:db}) describes the competition of two effects:
contraction and production. As a result of the balance between these
effects, a {\em steady state} sets in after sufficiently long time for
the global distribution of the B-particles [see
Figs.~\ref{fig:sim}(c) and \ref{fig:sim}(d)].  This steady state is synchronized to the
flow, i.e., it takes over the time dependence of the flow and in general follows the
hydrodynamical time dependence manifested in the parameter $c$.  In the case when the time dependence of parameter $c$ is weak,
the steady state value of the number of particles is
\begin{equation}
B^*=  \left( \frac{c v_r}{\lambda} \right)^{2-D} \sim
{\delta^*}^{2-D}.
\label{eqn:bs}
\end{equation}
The scaling is unique: $B^*$ is
proportional to the power $(2-D)<1$ of the reaction velocity.  This
means that the number $B^*$ of particles decreases by a factor smaller than the factor by which the
reaction velocity decreases.  For example, if $v_r \sim 10^{-4}$, the value of $B^*$ with $D=1.5$ is two orders of magnitude larger than that for the traditional active process where $D=1$. Equation~(\ref{eqn:bs}) is confirmed in
Figs.~\ref{fig:sim}(e) and \ref{fig:sim}(f) for inertial particles whose dynamics
possesses a chaotic attractor and transient chaos, respectively.  

\noindent{\it Remark I: Production vs diffusivity.}
Because the reaction front velocity $v_r$ is known\cite{Kol, Fischer} to be
proportional to the square root of the diffusion coefficient $\kappa$,
we obtain from~\eqref{eqn:bs}
\[ B^* \sim \kappa^{1-D/2}.\]
The amount of particles produced is proportional to the fractional
power ($1-D/2$) of the diffusion coefficient.  This relation for
diffusive particles has been derived using an Eulerian approach
in \cite{Tan,Wonhas}.  Our arguments
herewith imply that it is in fact valid for active inertial particles
as well, provided $v_r \sim \kappa^{1/2}.$

\noindent{\it Remark II: Dependence of production on resolution.}
Let us now consider the production term $P$, but for simplicity, in the
steady state.  Assume that this production is measured with a
resolution $\varepsilon$ {\em worse} than the crossover scale length, i.e.,
$\varepsilon >\delta^*$.  Since the production is proportional to the
perimeter length seen with the resolution used, we have
\begin{equation}
P(\varepsilon) \sim
{\varepsilon}^{1-D}.
\end{equation}
The exact amount of production $P(\delta^*)$ is, however, proportional
to ${\delta^*}^{1-D}$. The ratio of the observed, coarse-grained
amount of production to the exact one is thus
\begin{equation}
\frac{P(\varepsilon)}{P(\delta^*)} =
\left( \frac{\delta^*}{\varepsilon} \right)^{D-1}.
\end{equation}
By improving the resolution (decreasing $\varepsilon$ to $\delta^*$)
the ratio moves towards unity.  This dependence is not present at all
in the nonchaotic case where $D = 1$.  Therefore, we conclude that the
increase of productivity with increasing resolution observed earlier
in simulations of environmental problems\cite{Edouard,Mahad} is describable by the
equation derived in this paper.  Our results show that this effect is
present even when the description of the hydrodynamical flow field is
complete, in contrast to the similar effect reported before\cite{Edouard,
Mahad}, which may be due to incomplete knowledge of the flow field.
Although previous studies\cite{Edouard,Mahad} treat only noninertial particles,
it follows from our approach that this behavior must be present in the
inertial problem as well.

\noindent{\it Remark III: Enhancement factor.}
In a nonchaotic flow, the average width $\delta^*$ of B-particle bands
in the steady state is proportional to $v_r/\lambda$.  The enhancement
factor relative to the nonchaotic case is thus
\begin{equation}
\frac{B^* (D)}{B^*(D=1)} \cong \left( \frac{v_r}{\lambda} \right)^{1-D}. 
\label{eqn:ef}
\end{equation}
Since $\delta^*$ is typically much smaller than the characteristic
length scale of the flow (chosen here to be unity), there is always a
considerable enhancement due to the chaoticity of the advection
dynamics (recall $1-D < 0$)\cite{footnote2}.

We emphasize that these results are independent of the nature of
reaction (autocatalytic, bistable, excitable, etc.), that is, whenever
the product is distributed in bands along filaments of a fractal set,
Eqs. (\ref{eqn:db})--(\ref{eqn:ef}) hold. 

\begin{figure}
\begin{center}
  \includegraphics{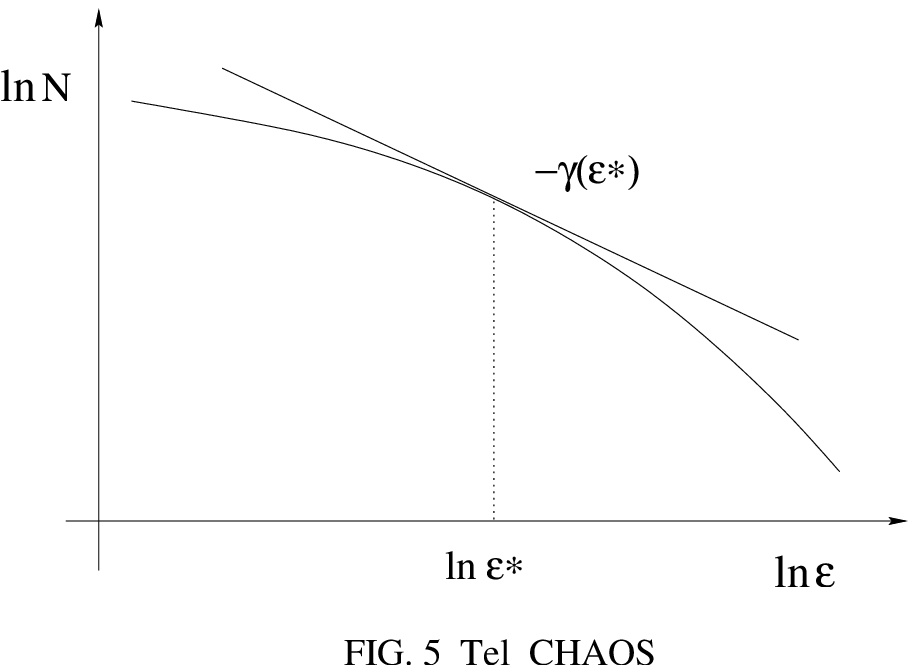}
\end{center}
\caption{The definition of the exponent
$\gamma(\varepsilon^*)$ for a given resolution $\varepsilon^*$. \label{fig:cart}}
\end{figure}

\noindent{\it Remark IV: Effects of nonfractal filaments.}
The existence of a well-defined fractal
scaling over decades of resolution is actually not necessary for our theory to
work.  For systems close to the steady state, our approach only
requires that $N(\delta)$ is a power law of $\delta$ at the length
scale around $\delta^*$.  This is important because many filamental
structures observed in environmental processes do not present a clear
scaling over decades of resolution.  This is the case, for example,
for the plankton growth\cite{Martin}, or for the deactivation process
of the ClO-rich polar air by the NO$_2$-rich air in the mid-northern
latitudes, which can suppress the depletion of ozone\cite{Wonhas}.
Our treatment can be carried out in this case with an exponent
$\gamma(\varepsilon=\delta^*)$ replacing $D$, where $\gamma(\varepsilon)$ is
defined as the slope of the $\ln N(\varepsilon)$ vs $\ln
(1/\varepsilon)$ curve (see Fig.~\ref{fig:cart}):
\[ \gamma(\varepsilon) = - \frac{d \ln N(\varepsilon)}{d \ln \varepsilon}. \]
The exponent $\gamma$ is a quantity which would be the fractal
dimension $D$ if the exact scaling $N(\varepsilon) \sim
\varepsilon^{-D}$ holds with a constant $D$.  Equations~(\ref{eqn:db})--(\ref{eqn:ef}) remain valid if we substitute
\[ D \rightarrow \gamma(\delta^*). \]
Note, however, that, in contrast to the case of a clear fractality,
the use of our theory in this case requires the knowledge of the
steady state width $\delta^*$ and the assumption that $B$ is close to
its steady state value $B^*$ so that $\gamma(\delta^{*})$ stays approximately constant over time.  We observe that the exponent $\gamma$, relevant for reactions, can be different from the exact dimension of an
underlying fractal set, even when such a quantity exists (see
Appendix \ref{appx:B}).

\noindent{\it Conclusions:}
Starting from a particle-based ``microscopic'' picture, we derived, by
applying elementary rules of the fractal geometry, a novel type of
rate equation in which the production term does not follow the
principle of ``mass-action'' \cite{Reif} well known from
thermodynamics. In fact, these processes are much further away from
thermal equilibrium than traditional reactions since they do not fill
the configuration (or phase) space.  The problem treated here provides
a clean example of a feature we believe to be general: whenever a
transport process is concentrated on a fractal set in the
configuration space, the corresponding {\em transport equation
deviates substantially} from the one known from irreversible
thermodynamics.

\appendix

\section{Equation of motion for inertial particles\label{appx:A}}

For small spherical particles of finite size, the particle velocity ${\bf v}
= d{\bf r}/dt$ (${\bf r}$ is the position of the particle) typically
differs from the fluid velocity ${\bf u}$.  The equation of motion is
given by Newton's second law: the force causing the relative
acceleration $d {\bf v}/dt - d {\bf u}/dt$ [$d {\bf u}/d t \equiv
\partial {\bf u}/\partial t + ({\bf u} \cdot \nabla) {\bf u}$] between
the particle and the fluid is due to the viscous friction and, in the
gravitational field, due to buoyancy.  The former, the so-called Stokes
drag, is proportional to the velocity difference ${\bf v}-{\bf u}$,
and vanishes for pointlike particles.  The latter is proportional to
the density difference $\varrho_{\mbox{\scriptsize
particle}}-\varrho_{\mbox{\scriptsize fluid}}$. The dimensionless form
of the equation of motion reads\cite{Max1,Max2,Balk,footnote3}:
\begin{equation}
\frac{d {\bf v}}{d t} - \alpha
\frac{d {\bf u}}{d t} = -
\frac{{\bf v}-{\bf u}}{St} - w {\bf n}.
\label{em}
\end{equation}
Here $St>0$ is the Stokes number, the dimensionless decay time due to
the Stokes drag, $w$ (positive for heavy particles) is the
dimensionless buoyancy force acting in the vertical direction, and
${\bf n}$ denotes the vertical normal vector pointing upward.  The
coefficient $\alpha>0$ expresses the fact that a finite size particle
brings into motion a certain amount of fluid proportional to its
volume.  The noninertial particle dynamics is recovered when the
particle radius vanishes, which corresponds to the limit $St
\rightarrow 0$.  In this limit, the advection dynamics is governed by
\[ {\bf v} = \frac{d \bf r}{dt} = {\bf u}({\bf r}, t). \] 
The general inertial dynamics (\ref{em}) possesses a four-dimensional
phase space $(x,y,v_x,v_y)$ even for planar stationary flows, whereas for the
non-inertial particle dynamics the phase space is two dimensional.
The inertial dynamics is dissipative, even in incompressible flows,
and the phase space volume contracts at the rate $-2/St$, which is
always negative, in contrast to the noninertial case which is volume
preserving.  The fractal object in the full phase space must have a
dimension less than $2$, if one wishes to keep its fractality in the
projection onto the configuration space of the flow.

\section{Difference between $\gamma(\varepsilon)$ and fractal dimension\label{appx:B}}

\begin{figure}
\begin{center}
  \includegraphics{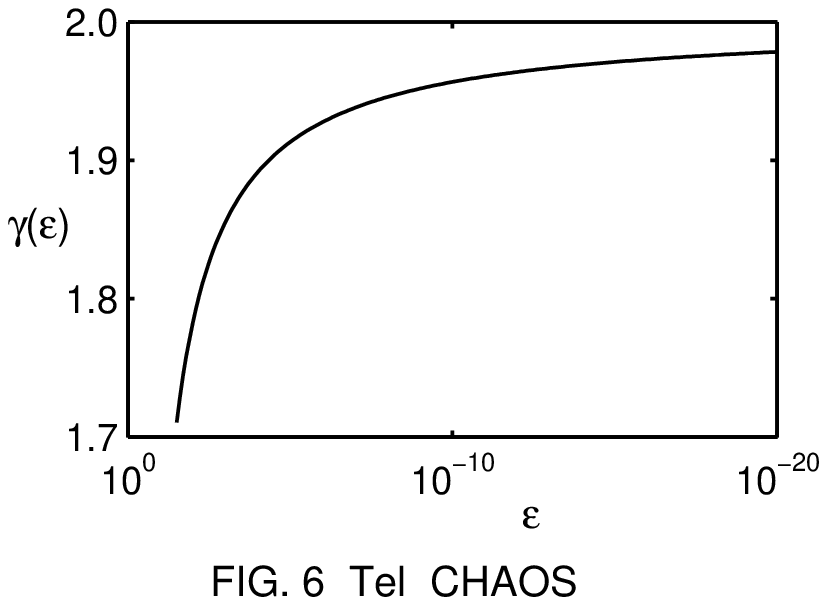}
\end{center}
\caption{The exponent $\gamma$ as a function of the observational scale
$\varepsilon$ for the filamental fractal constructed in
Appendix \ref{appx:B}, which models fractals observed in typical nonhyperbolic
systems. \label{fig:deff}}
\end{figure}

An illustrative example where the exponent $\gamma(\varepsilon)$
differs from the fractal dimension is the case of point particles
advected by a flow in which chaotic and regular motions coexist
(the so-called nonhyperbolic dynamics).  A transverse intersection
between a line and the filamental fractal associated with such a chaotic
motion can be modeled by a Cantor set constructed as follows.  In the
first step, an interval is removed from the middle of the unit
interval.  An interval is removed from each of the remaining two
intervals in the second step.  The third step removes intervals from
the middle of all remaining interval, and so on.  In contrast to the
hyperbolic chaotic systems, the relative size of removed intervals at
each step is not constant, but, for example, is inversely proportional
to the number of steps in the construction\cite{Lau}.  In this case,
the limit set in the two-dimensional space is a fractal set of
dimension two.  However, the exponent $\gamma$ at resolution
$\varepsilon$ is smaller than 2 and can be approximated by
\[ \gamma(\varepsilon) \approx 2 - \frac{1}{\ln(1/\varepsilon)}. \]
The exponent $\gamma$ converges very slowly to the exact dimension 2
and it is quite different from the limiting value, even for
unrealistically small scales, as shown in Fig.~\ref{fig:deff}.
A similar behavior for the exponent $\gamma$ is also expected in the
presence of small dissipation due to inertial
effects\cite{Motter1,Motter2}.  If a reactive process takes place in
such a nonhyperbolic advection dynamics, $D$ in (\ref{eqn:db})--(\ref{eqn:ef}) must
be replaced by the exponent $\gamma(\varepsilon = \delta^*)$ in a self-consistent manner.

\begin{acknowledgments}
This work was supported by the Hungarian Science Foundation (OKTA T032423) and the MTA/OTKA/NSF Fund (Project No. Int. 526).  AEM and CG were supported by FAPESP, and CG was additionally supported by CNPq.  ZT was supported by the DOE under contract No. W-7405-ENG-36.  We thank Max-Planck-Institut f\"ur Physik Komplexer Systeme, Dresden, Germany where some of the key ideas of this work were discussed.
\end{acknowledgments}

%\bibliographystyle{apsrev}
%\bibliography{activechaos}

\begin{thebibliography}{50}
\bibitem{footnote} In systems with multifractality, we must replace $D$
with the information dimension $D_1$, because the advected particles
only cover typical points of the fractal set.
\bibitem{footnote2} After normalizing by the time scale of the flow, there are four relevant dimensionless parameters that are independent in the problem: the Stokes time ($St$), the Lagrangian time scale ($\sim 1/\lambda$), the diffusion time scale ($\sim 1/\kappa$), and the reaction time scale ($\sim 1/v_r)$.  The diffusion time scale is much longer than the others, but it does not appear explicitly in our approach (only via $v_r$).  In terms of the other time scales, the condition for strong enhancement is the slow front propagation on the Lagrangian time scale.
\bibitem{footnote3} The Basset history term is neglected here since it is expected to
be important only if the particle returns to previously visited cites,
which is not the case for rising or descending particles
under gravitational force ($-w \textbf{n}$) [see E. E. Michaelides, J. Fluids
Eng. \textbf{119}, 233 (1997)].
\end{thebibliography}

%%%%%%%%%%%%%%%%%%%%%%%%%%%%%%%%%%%%%%%%%%%%%%%%%%%%%%%%%%%%%
% Figure captions
%
%\newpage

\vspace{5mm}
\newcounter{fig}
%\refstepcounter{fig}
%\begin{center}
%  \includegraphics{fig1.eps}
%\end{center}
%\noindent FIG.\,\,\thefig: Filamentation in a phytoplankton bloom in Norwegian Sea. (Provided by the \mbox{SeaWiFS} Project, NASA/Goddard Space Flight Center, and ORBIMAGE, URL: {\tt http://visibleearth.nasa.gov/cgi-bin/viewrecord?5278}) \label{fig:real}

%\vspace{5mm}
%\refstepcounter{fig}
%\begin{center}
%  \includegraphics{fig2.eps}
%\end{center}
%\noindent FIG.\,\,\thefig: Local reaction-advection dynamics on a filamental segment. \label{fig:react}

\vspace{5mm}
\refstepcounter{fig}

\vspace{5mm}
\refstepcounter{fig}
%\begin{center}
%  \includegraphics{fig4.eps}
%\end{center}
%\noindent FIG.\,\,\thefig: Schematic diagram illustrating the
%$\varepsilon$-dependence of the perimeter length ${\cal L}$ and the area
%${\cal A}$ of a filamental fractal, when observed with resolution
%$\varepsilon$. \label{fig:cover}

\vspace{5mm}
\refstepcounter{fig}
%\begin{center}
%  \includegraphics{fig5.eps}
%\end{center}
%\noindent FIG.\,\,\thefig: The definition of the exponent
%$\gamma(\varepsilon^*)$ for a given resolution $\varepsilon^*$. \label{fig:cart}

\vspace{5mm}
\refstepcounter{fig}
%\begin{center}
%  \includegraphics{fig6.eps}
%\end{center}
%\noindent FIG.\,\,\thefig: The exponent $\gamma$ as a function of the observational scale
%$\varepsilon$ for the filamental fractal constructed in
%Appendix \ref{appx:B}, which models fractals observed in typical nonhyperbolic
%systems. \label{fig:deff}
%}
%\end{figure}

\end{document}